\documentstyle{article}
\renewcommand{\theequation}{\thesection.\arabic{equation}}
\newcounter{hran} \renewcommand{\thehran}{\thesection.\arabic{hran}}

\def\bmini{\setcounter{hran}{\value{equation}}
  \refstepcounter{hran}\setcounter{equation}{0}
  \renewcommand{\theequation}{\thehran\alph{equation}}\begin{eqnarray}}

\def\bminiG#1{\setcounter{hran}{\value{equation}}
\refstepcounter{hran}\setcounter{equation}{-1}
\renewcommand{\theequation}{\thehran\alph{equation}}
\refstepcounter{equation}\label{#1}\begin{eqnarray}}

\renewcommand{\theequation}{\thesection.\arabic{equation}}

\def\bmini{\setcounter{hran}{\value{equation}}
  \refstepcounter{hran}\setcounter{equation}{0}
  \renewcommand{\theequation}{\thehran\alph{equation}}\begin{eqnarray}}

\def\bminiG#1{\setcounter{hran}{\value{equation}}
\refstepcounter{hran}\setcounter{equation}{-1}
\renewcommand{\theequation}{\thehran\alph{equation}}
\refstepcounter{equation}\label{#1}\begin{eqnarray}}

\def\emini{\end{eqnarray}\relax\setcounter{equation}
{\value{hran}}\renewcommand{\theequation}{\thesection.\arabic{equation}}}

\newcommand{\bal}{\begin{array}{ll}} 
\newcommand{\eal}{\end{array}}

\def\1{{\rm 1 \kern -.10cm I \kern .14cm}} \def\R{{\rm R \kern -.28cm I
\kern .19cm}}
\def\emini{\end{eqnarray}\relax\setcounter{equation
}{\value{hran}}\renewcommand{\theequation}{\thesection.\arabic{equation}}}

\newcommand{\be}{\begin{equation}}
\newcommand{\ee}{\end{equation}}
\def\bea{\begin{eqnarray}}
\def\eea{\end{eqnarray}}

\newcommand{\barre}[1]{%
	\setbox1=\hbox{$#1$} \dimen2=\ht1 \dimen3=\dp1 \dimen4=\wd1
	\setbox2=\hbox{\sl /}
	\dimen1=\wd1 \advance\dimen1 by -\wd2 \divide\dimen1 by 2
	\advance\dimen1 by \wd2 \advance\dimen1 by 0.4pt
	\setbox3=\hbox to \wd1{\hss \box1 \kern -\dimen1 \box2\hss}
	\ht3=\dimen2 \dp3=\dimen3 \wd3=\dimen4
	\box3
	}

\def\1{{\rm 1 \kern -.10cm I \kern .14cm}} \def\R{{\rm R \kern -.28cm I
\kern .19cm}}
\begin{document}
\begin{titlepage}
\begin{flushright}    UFIFT-HEP-98-38 \\ 
\end{flushright}
\vskip 1cm
\centerline{\LARGE{\bf {Anomalous $U(1)$, holomorphy,}}}
\centerline{\LARGE{\bf {supersymmetry breaking and dilaton stabilization}}}
\vskip 1.5cm
\centerline{\bf Nikolaos Irges \footnote{e-mail address:
irges@phys.ufl.edu}}

\vskip .7cm
\centerline{\em Institute for Fundamental Theory,}
\centerline{\em Department of Physics, University of Florida}
\centerline{\em Gainesville FL 32611, USA}
\vskip .2cm
\vskip 1.5cm
\centerline{\bf {Abstract}}
\vskip .5cm

We argue that in certain models with family symmetries the implementation of
the alignment mechanism for the supression of the flavor changing
neutral currents requires mass matrices with
holomorphic zeros in the down quark sector.
Holomorphic zeros typically open flat directions 
that potentially spoil the uniqueness of the supersymmetric vacuum. 
We then present an anomalous $U(1)$ model
without holomorphic zeros in the quark sector
that can reproduce the fermion mass hierarchies, provided that 
$\tan{\beta}$ is of order one. To avoid
undesired flavor changing neutral currents
we propose a supersymmetry breaking mechanism and a dilaton
stabilization scenario that result in degenerate squarks at $M\sim
M_{GUT}$ and a calculable low energy spectrum. 
We present the numerical predictions of this 
model for the Higgs mass for different values of $M$ and $\tan{\beta}$.         

\vfill

\end{titlepage}

\section{Introduction}

Recently\cite{EIR2}, it was shown that in models of fermion masses 
with family symmetries,
the assumption of no holomorphic zeros 
in the quark sector of the superpotential
uniquely determines the form of the $U(1)$ that reproduces the data, 
to 
$Y_X\equiv X+Y_F\equiv X+Y^{(1)}+Y^{(2)}$, with
\be Y^{(1)}={1\over 5}(2Y+V)(2,-1,-1) \;\;\; {\rm and}
\;\;\; Y^{(2)}=-{1\over 2}(V+3V')(1,0,-1),\ee  
where the vectors $(2,-1,-1)$, $(1,0,-1)$ show the family dependence, 
$Y$ is hypercharge
and $V,V'$ are the other two anomaly free
$U(1)$'s in the ${\bf 27}$ of $E_6$. $X$ is a family 
independent, anomalous trace which had to be added to implement
the correct intrafamily hierarchy \cite{EIR2},\cite{ILR}. 
The matter content is 3 families of fields filling out 
the ${\bf 27}$ of $E_6$, except the singlet outside $SO(10)$. 
One interesting prediction of this model is
the neutrino mixing matrix that can be conveniently expressed
in terms of the Cabbibo angle $\lambda_c$, as:
\be \pmatrix{1 & \lambda _c^3 & \lambda _c^3\cr 
\lambda _c^3 & 1 & 1\cr \lambda _c^3 & 1 & 1},\ee
consistent with the atmospheric neutrino data coming from SuperK and
the non-adiabatic MSW solution to the solar neutrino flux anomaly.
Any complete supersymmetric 
model of masses however, has to also give an explanation for
the suppression of flavor changing neutral current
(fcnc) processes, which generically get large contribution in the presence 
of an anomalous $U(1)$\cite{Ross}, due to the nonzero $D$-term associated with it.
\par A possibility of solving this problem is when there is 
alignment between the up and down sectors
\cite{NS}. 
One striking property of this mechanism is that there is no need for
degenerate squarks after supersymmetry breaking,
to suppress fcnc. We will show soon that
alignment can take place only if the mass matrices in the up and down quark
sectors have the form:
\be Y_u^{\rm align}=\pmatrix {\lambda _c^8 & 
\lambda _c^5 & \lambda _c^3 \cr * & \lambda _c^4 & \lambda _c^2 \cr 
* & * & 1} \;\;\;\; {\rm and} \;\;\;\; Y_d^{\rm align}=
\pmatrix {\lambda _c^4 & 0 & \lambda _c^3 \cr 0 & \lambda _c^2 & \lambda _c^2 \cr 
0 & 0 & 1}  \label{eq:alignmatrix},\ee
where $*$ means an entry which is a holomorphic zero or not. 
Even though this is a possible solution to the fcnc problem, 
we choose not to favor it because holomorphic zeros tend to 
destabilize the vacuum, as was argued in \cite{IL}.      
\par In the case of absence of holomorphic zeros from the
quark sector of the superpotential, the contributions of the up and
down sectors to the CKM matrix is the same, alignment  
does not take place and therefore to suppress
fcnc, a supersymmetry mechanism which generates degenerate squarks is
necessary. This leads us to the scenario of dilaton
dominated supersymmetry breaking, which is a mechanism 
that can give degenerate squark masses at the scale $M\sim M_{GUT}$.
The main difficulties of constructing a predictive model of this kind is
to solve the problem of dilaton stabilization and to give an explanation for
why the $D$-term contributions to the soft masses is suppressed.
\par In section II, we show that the most general form of matrices
that can implement the alignment mechanism is that of \ref{eq:alignmatrix},
which contains undesired, in our point of view, holomorphic zeros. 
In section III, we turn into models with minimum number 
of holomorphic zeros and we show that it is possible to
stabilize the dilaton via non perturbative corrections to the 
K$\ddot{\rm a}$hler potential and at the same time suppress the
$D$-term contributions to the soft masses, so that the (family
universal) contribution of the dilaton $F$-term dominates the squark masses.
We give an explicit example of a globally supersymmetric 
model that has these properties
and finally we argue that supergravity corrections do not modify 
this picture. 

\section{Holomorphic zeros and alignment}

We consider models with $N+1$ family $U(1)$ symmetries. 
If $N=0$, there is only one such $U(1)$, which 
has to be anomalous, as mentioned in the introduction.  
Even though we do not have a proof for it, cancelation of anomalies
containing hypercharge, strongly indicate that it is the trace of 
the $U(1)$ that carries all of the anomaly, leaving its traceless
part anomaly free. If $N\ge 1$, for the same reason, without loss 
of generality, we will assume that all of the anomaly is contained
in a family independent $U(1)$, leaving the other $N$ traceless
and anomaly free. In both cases we will call the anomalous part $X$.
Also, in a model with $N+1$ $U(1)$'s, we assume $N+1$ order parameter
fields $\theta_{\alpha}$, ($\alpha =0,\cdots ,N$), 
that take vevs and break those $U(1)$'s. 
A gauge invariant term in the superpotential has the form: 
\be {\Bigl({\theta _0\over M}\Bigr)}^{n_0^{i_1i_2i_3...i_I}}
\cdots {\Bigl({\theta _{N}\over M}\Bigr)}^{n_{N}^{i_1i_2i_3...i_I}}
{\bf I}_{i_1i_2i_3...i_I}.\ee
We have displayed in the 
invariant ${\bf I}$ and the exponents the family indices explicitly. 
The invariance of this term under the $U(1)$'s 
allows us to compute the powers $n_{\alpha}^{i_1...i_I}$
from the matrix equation
${\bf n}=-{\cal A}^{-1}{\bf Y}^{(I)}$, or explicitly:
\be \pmatrix {n_0^{i_1i_2i_3...i_I}
\cr . \cr . \cr . \cr n_N^{i_1i_2i_3...i_I} }=-{\cal A}^{-1}
\pmatrix {X({\bf I}_{i_1i_2i_3...i_I})
\cr Y^{(1)}({\bf I}_{i_1i_2i_3...i_I})\cr . \cr . \cr . \cr 
Y^{(N)}({\bf I}_{i_1i_2i_3...i_I})}, \ee
where we have introduced the following notation:
${\bf n}$ is an $(N+1)\times 1$ column vector with 
the powers of the $\theta $ fields, 
${\bf Y}^{(I)}$ is an $(N+1)\times 1$ column vector 
with the charges of the standard model invariant 
${\bf I}_{i_1i_2i_3...i_I}$ under the 
$N+1$ $U(1)$'s,
${\cal A}$ is the $(N+1)\times (N+1)$ matrix whose first
row is the $X$ charges of $\theta_{\alpha}$ and its 
$a$'th row ($a=1,\cdots ,N$) is the $Y^{(a)}$ charges of $\theta_{\alpha}$. 
${\cal A}^{-1}$ is assumed to be of such form so that
all the elements of its first column 
are 1. This means that all the 
$\theta $ fields ($N+1$ of them) take vacuum
expectation values at the same scale. 
\footnote{We will keep this assumption until the end of
this paper, because it makes the discussion on mass matrices
easier. We could relax this assumption and still arrive at the same
conclusions.}
We denote the
common vev $<\theta_{\alpha}>/M$ by $\lambda _c$
and later identify it with the Cabbibo angle. 
We set the notation for the Abelian charges 
of the observed quarks under the $a$'th non-anomalous $U(1)$: 
\vskip 0.2cm
\hskip 2cm
\begin{center}
\begin{tabular}{|c|c|c|c|}
\hline 
$ $ & ${\rm { 1st \; {\rm family}}} $ & $ 2{\rm nd} \; 
{\rm family}$ & $3{\rm rd} \; {\rm family}$  \\
\hline \hline   
${\bf Q}$ & $q_1^{[a]}$ & $q_2^{[a]}$ & $-q_1^{[a]}-q_2^{[a]}$\\ \hline
${\bf \overline u}$ & $u_1^{[a]}$ & 
$u_2^{[a]}$ & $-u_1^{[a]}-u_2^{[a]}$\\ \hline                                 
${\bf \overline d}$ & $d_1^{[a]}$ & 
$d_2^{[a]}$ & $-d_1^{[a]}-d_2^{[a]}$  \\ \hline
\end{tabular}
\end{center}
\vskip 0.3cm
In the case of a
single, anomalous $U(1)$, 
the above table gives the charges of its traceless part. 
\par It is useful to introduce the quantities
\be Q_{12}^{[a]}\equiv 2q_1^{[a]}+q_2^{[a]} \;\;\;\; 
{\rm and} \;\;\;\; Q_{21}^{[a]}\equiv 2q_2^{[a]}+q_1^{[a]}\label{eq:Ch1}, \ee
\be U_{12}^{[a]}\equiv 2u_1^{[a]}+u_2^{[a]} \;\;\;\; 
{\rm and} \;\;\;\; U_{21}^{[a]}\equiv 2u_2^{[a]}+u_1^{[a]}\label{eq:Ch2}, \ee
\be D_{12}^{[a]}\equiv 2d_1^{[a]}+d_2^{[a]} \;\;\;\; {\rm and} 
\;\;\;\; D_{21}^{[a]}\equiv 2d_2^{[a]}+d_1^{[a]}.\label{eq:Ch3} \ee
In the up and down quark sectors we get the Yukawa matrices 
\be Y_{u}=\lambda _c^{N^{[u]}}
\pmatrix {\lambda_c^{M+K} & \lambda_c^{P+K} & \lambda_c^{0+K} \cr 
              \lambda_c^{M+L} & \lambda_c^{P+L} & \lambda_c^{0+L} \cr
              \lambda_c^{M+0} & \lambda_c^{P+0} & \lambda_c^{0+0} \cr
} \label{eq:matintup} \;\; {\rm and} \;\; 
Y_{d}=\lambda _c^{N^{[d]}}
\pmatrix {\lambda_c^{R+K} & \lambda_c^{T+K} & \lambda_c^{0+K} \cr 
              \lambda_c^{R+L} & \lambda_c^{T+L} & \lambda_c^{0+L} \cr
              \lambda_c^{R+0} & \lambda_c^{T+0} & \lambda_c^{0+0} \cr
}\label{eq:sumrule}.\ee
$N^{[u]}$, $N^{[d]}$ are defined as the total powers appearing 
at the 33 position of the corresponding 
mass matrix which we always pull out front.   
Also,  
\be \pmatrix {K \cr L \cr M \cr P \cr R \cr T} =
\sum _{{\alpha}}\pmatrix {K_{\alpha} \cr L_{\alpha} \cr M_{\alpha} 
\cr P_{\alpha} \cr R_{\alpha} \cr T_{\alpha}}, \;\;\; {\rm where} \;\;\;
\pmatrix {K_{\alpha} \cr L_{\alpha} \cr M_{\alpha} 
\cr P_{\alpha} \cr R_{\alpha} \cr T_{\alpha}}=
-{({\cal A}^{-1})}_{{\alpha}{\beta}} \pmatrix 
{Q_{12}^{[{\beta}]} \cr Q_{21}^{[{\beta}]} \cr 
U_{12}^{[{\beta}]} \cr U_{21}^{[{\beta}]} 
\cr D_{12}^{[{\beta}]} \cr D_{21}^{[{\beta}]}} \label{eq:integers}.\ee 
As usual, ${\alpha},{\beta}=0,...,N$ and summation over ${\beta}$ is
implied. 
\par The quark matrices are diagonalized as 
\be Y_{u}={V}^{u^\dag }_L M_{u}V^{u}_R,\;\;\;\;\;\; 
Y_{d}={V}^{d^\dag }_L M_{d}V^{d}_R.\ee
From the above and \ref{eq:sumrule}, 
which imply $K=3$, $L=2$ with $K_{\alpha},L_{\alpha}\ge 0$, 
we see that our notation for 
$\lambda_c$ is fully justified.
Similarly, for the squarks we have mass matrices 
associated with the soft terms 
${\tilde m}_{ij}^2{\bf \tilde q}^{*}_i{\bf \tilde q}_j$,
which can be diagonalized as
\be {\tilde Y}^{u,d}_{LL}={\tilde V}^{{u,d}^\dag }_L 
{\tilde M}^{u,d}_{LL}{\tilde V}^{u,d}_L, \;\; 
{\tilde Y}^{u}_{RR}={\tilde V}^{u^\dag }_R 
{\tilde M}^{u}_{RR}{\tilde V}^{u}_R \;\; , \;\;
{\tilde Y}^{d}_{LL}={\tilde V}^{d^\dag }_L 
{\tilde M}^{d}_{RR}{\tilde V}^{d}_L .\ee
The most stringent experimental limits on the entries of these matrices 
(coming primarily from the neutral meson mixing data) 
are \cite{NS}: 
\begin{eqnarray}(K_L^d)_{12}=(V_L^d{\tilde V}_L^{d^\dag })_{12}=
\lambda _c^{m^{L_d}_{12}}\leq \lambda _c^3, \;\;\;\; {\rm where} \cr \;\; 
(K_L^d)_{12}=max\Bigl[({\tilde V}_L^{d^\dag })_{12},
({V}_L^{d^\dag })_{12},({V}_L^{d^\dag })_{13}\cdot ({\tilde V}_L^{d^\dag
})_{32}\Bigr], \label{eq:fcnc1} \end{eqnarray}
\begin{eqnarray}(K_R^d)_{12}=(V_R^d{\tilde V}_R^{d^\dag })_{12}=
\lambda _c^{m^{R_d}_{12}}\leq \lambda _c^3, \;\;\;\; {\rm where} \cr \;\; 
(K_R^d)_{12}=max\Bigl[({\tilde V}_R^{d^\dag })_{12},
({V}_R^{d^\dag })_{12},({V}_R^{d^\dag })_{13}\cdot 
({\tilde V}_R^{d^\dag })_{32}\Bigr],\label{eq:fcnc2}\end{eqnarray}
If there are no supersymmetric zeros in the 
mass matrices, we can explicitly diagonalize them. We obtain
\be 
{V}^{u}_L=\pmatrix {1 & \lambda_c^{<K_{\alpha}-L_{\alpha}>} & 
\lambda_c^{<K_{\alpha}>} \cr 
\lambda_c^{<K_{\alpha}-L_{\alpha}>} & 1 & \lambda_c^{<L_{\alpha}>}
\cr \lambda_c^{<K_{\alpha}>} & \lambda_c^{<L_{\alpha}>} & 1 }, \;\; 
{V}^{u}_R=\pmatrix {1 & \lambda_c^{<M_{\alpha}-P_{\alpha}>} & 
\lambda_c^{<M_{\alpha}>} \cr 
\lambda_c^{<M_{\alpha}-P_{\alpha}>} & 1 & \lambda_c^{<P_{\alpha}>}
\cr \lambda_c^{<M_{\alpha}>} & \lambda_c^{<P_{\alpha}>} & 1 }
\ee
\be 
{V}^{d}_L=\pmatrix {1 & \lambda_c^{<K_{\alpha}-L_{\alpha}>} & 
\lambda_c^{<K_{\alpha}>} \cr 
\lambda_c^{<K_{\alpha}-L_{\alpha}>} & 1 & \lambda_c^{<L_{\alpha}>}
\cr \lambda_c^{<K_{\alpha}>} & \lambda_c^{<L_{\alpha}>} & 1 }, \;\; 
{V}^{d}_R=\pmatrix {1 & \lambda_c^{<R_{\alpha}-T_{\alpha}>} & 
\lambda_c^{<R_{\alpha}>} \cr 
\lambda_c^{<R_{\alpha}-T_{\alpha}>} & 1 & \lambda_c^{<T_{\alpha}>}
\cr \lambda_c^{<R_{\alpha}>} & \lambda_c^{<T_{\alpha}>} & 1 }.
\ee
The $<.>$ symbol means summation over $\alpha$. To obtain the matrices
${\tilde V}$ in the squark sector, all we have to do is to replace $<.>$
by $<|.|>$, where $|.|$ means absolute value. 
We can now compute for example 
\begin{eqnarray} {m^{R_d}_{12}}=min\Bigl[<{(R_{\alpha}-T_{\alpha})}>,
<{|R_{\alpha}-T_{\alpha}|}>,
<{(R_{\alpha}+|T_{\alpha}|)}>\Bigr]
= \cr =<{(R_{\alpha}-T_{\alpha})}>=
<{R_{\alpha}}>-<{T_{\alpha}}>=1-0=1.\label{eq:minimum} \end{eqnarray}
By looking at \ref{eq:fcnc2}, we see that the fcnc constraints are not satisfied. 
In fact, in order to have a hope for satisfying the alignment constraints,
$Y_{d_{12}}$ has to be a holomorphic
zero, so that in \ref{eq:minimum}, the factor $<{(R_{\alpha}-T_{\alpha})}>$ 
that causes the misalignment is missing. To see this, notice that in
order to reproduce the correct $CKM$ matrix and  
intrafamily hierarchy, we have to have
$K_{\alpha},L_{\alpha}>0$ and $N^{[d]}_{\alpha}>0$ respectively.
Similarly, from \ref{eq:fcnc1}, we conclude that at the same time 
$Y_{d_{21}}$ has also to be a zero.
But if these are zeros,
then also $Y_{d_{32}}$ and $Y_{d_{31}}$ have to be 
holomorphic zeros as well, as the sum rules \ref{eq:sumrule} 
indicate. This is the minimum
number of supersymmetric zeros in the down sector 
and it is also the maximum since neither the diagonal elements, 
nor $Y_{d_{13}}$ and $Y_{d_{23}}$
can be zeros if $Y_{d}$ should give the desired 
mass ratios and mixings. We have therefore showed that there is a unique
$Y_{d}$ 
compatible with the ``alignment scenario'' of suppressing fcnc. 
On the other hand, $Y_{u}$ 
is fixed except the elements $(21)$, $(31)$ and $(32)$. These are either
supersymmmetric zeros or not and we 
thus verified the matrix forms in \ref{eq:alignmatrix}. 
\par As we mentioned 
in the introduction, vacuum stability arguments suggest that we prefer
models with the least number of holomorphic zeros and therefore we now turn 
to the analysis of those models.  

\section{No holomorphic zeros, dilaton stabilization and supersymmetry breaking}

For a supersymmetry breaking mechanism 
with gaugino condensation in the hidden sector\cite{nilles},
in the presence of an anomalous
$U(1)$ \cite{BD} and where the dilaton dominates,
it has been shown \cite{dilaton} that the dilaton can be stabilized
with a weakly coupled K$\ddot{\rm a}$hler potential, so that its
second derivative $K_2$ at the minimum of the scalar potential is very
small. In this scheme, however,  
the $D$-term contribution to the soft 
masses is generically rather large \cite{BDi3}. 
The purpose of this section is to propose a
mechanism which stabilizes the dilaton at a realistic
value and at the same time 
suppresses the $D$-term relatively to the dilaton $F$-term.
The
mechanism that does both of the above mentioned things simultaneously,
as far as we know, is new. 
To achieve this, we will assume a K$\ddot{\rm a}$hler potential which
is a combination of the one proposed in \cite{dilaton} and of a
similar one to the one used in \cite{BDi3}. 
We will see that it will be necessary to have a $K_2$ of order one in
our scheme.   
\par Let us first construct a model with no holomorphic zeros
in the quark sector and a single $U(1)$. 
There is, in this case only one $\theta $-field ($\theta _0\equiv \theta$). 
We normalize the
charge of $\theta $ to be 1, which gives ${\cal A}=1$.  
Then, noticing that the data implies 
$K=3, L=2, M=5, P=2, R=1$ and $T=0$, equations 
\ref{eq:Ch1} to \ref{eq:integers}, give
\be \pmatrix {q_1 \cr q_2} = \pmatrix {{2\over 3} 
& -{1\over 3} \cr -{1\over 3} & {2\over 3} }  
\pmatrix {-3 \cr -2} = \pmatrix {-4/3 \cr -1/3},\ee
\be \pmatrix {u_1 \cr u_2} = \pmatrix {{2\over 3} 
& -{1\over 3} \cr -{1\over 3} & {2\over 3} }  
\pmatrix {-5 \cr -2} = \pmatrix {-8/3 \cr 1/3},\ee
\be \pmatrix {d_1 \cr d_2} = \pmatrix {{2\over 3} 
& -{1\over 3} \cr -{1\over 3} & {2\over 3} }  
\pmatrix {-1 \cr 0} = \pmatrix {-2/3 \cr 1/3}.\ee
We summarize in the following table the traceless part of our $U(1)$:
\vskip 0.3cm
\hskip 2cm
\begin{center}
\begin{tabular}{|c|c|c|c|}
\hline 
$  $ & $ {\bf Q} $ & $ {\bf \overline u} $ & $ {\bf \overline d} $  \\
\hline \hline   
$  $ & $ \pmatrix {-4/3 \cr -1/3 \cr 5/3} $ & $ 
\pmatrix {-8/3 \cr 1/3 \cr 7/3} $ & $ \pmatrix {-2/3 \cr 1/3 \cr 1/3} $  \\ \hline
\end{tabular}
\end{center}
\vskip 0.3cm
Therefore, the family dependent part of the 
symmetry acting upon the quark sector may be written as \cite{EIR2}
\be Y_F=B(2,-1,-1)-2\eta(1,0,-1)\label{eq:yf} ,\ee
where $B$ is baryon number, $\eta=1$ for both ${\bf Q}$ 
and ${\bf \overline u}$ and 
$\eta=0$ for ${\bf \overline d}$.
The next step is to supplement the quark sector by a
lepton sector and/or a vector like sector such that the
anomalies involving hypercharge that ought
to vanish, vanish. To our knowledge, this is possible to do in a
phenomenologically consistent way, only via
an $E_6$ embedding and specifically in the manner described in
\cite{EIR2}. We will not repeat this analysis here. We only mention
that the final result is the family $U(1)$ called $Y_X$ in the introduction.
If we gauge $X$, $Y^{(1)}$ and $Y^{(2)}$ separately,   
we have to supplement the visible sector of the model 
with a hidden sector as in \cite{ILR}, due to the 
appearance of the mixed anomaly $XY^{(1)}Y^{(2)}$,
which is nonzero and has to be canceled. We will take
this as a possible indication for the existence of a hidden sector.
The ratio of the $D$-term mass squared over the dilaton
$F$-term mass squared can be written in terms of the derivatives of
the K$\ddot{\rm a}$hler potential $K_i,\; i=1,2,3$ as \cite{dilaton}
\be {m_D^2\over m_{F_S}^2}={\bigl[-{K_3\over K_1}+({K_2\over K_1})^2\bigr]
\over {2K_2(1-4\pi ^2\delta_{GS}{K_2\over K_1}})},\ee
where $\delta_{GS}=Trace(X)/(192\pi^2)$.  
The task we have to carry out at this point is twofold. 
\par The first is
to propose a dilaton stabilization scenario with a specific  
K$\ddot{\rm a}$hler potential, which stabilizes the dilaton at a
value consistent with the value of the visible sector gauge couplings at the
unification point (in \cite{ILR}, the latter was computed to be
$\sim 3/2$). This can be achieved with a purely
weak coupling $K$ \cite{dilaton}:
\be K=-\log(2y)-{2s_0\over (2y)}+{{(4s_0^2+b)}\over {6(2y)^2}}\label{eq:Kweak},\ee
where $y$ is the real part of the dilaton field $S$ and $s_0$ and $b$
are numbers.
Then, the assumption of dilaton dominance, amounts to the scalar potential
of being simply
\be V={1\over K_2}|W_1|^2, \;\;\; {\rm with} \;\;\; 
K_2={1\over y^4}[(y-s_0)^2+{b\over 4}].\ee
$W_1$ is the first derivative of the 
(strong coupling) superpotential $W\sim e^{-mS}$ with respect
to $y$ ($m$ is a numerical constant that can be computed in a specific
model). Clearly, for $b\rightarrow 0$, if $y\rightarrow s_0$ then
$K_2\rightarrow 0$ and therefore $V$ approaches infinity. Due to the
exponentially decreasing form of $W$ (if $m>0$), the dilaton will roll down the
hill until it hits the bump located at $s_0$, provided $b$ is
small. In fact, the smaller $b$ is, the higher and narrower the bump
becomes. This will stabilize the dilaton at a value very close to
$s_0$. Let us denote that value by $y_0\equiv s_0-\alpha$, with
$\alpha$ being a small positive number.     
\par The second task is to explain why $m_D^2\ll m_{F_S}^2$, which 
is equivalent to 
$R_m\equiv -K_3/K_1+(K_2/K_1)^2\sim 0$. Using (\ref{eq:Kweak}), one can
verify that $K_2$ evaluated at the minimum of the potential is a very
small number: $K_2=1/(m^2s_0^2)$, which gives $R_m\sim s_0^2/b$, 
a rather large number. We see that a small $K_2$ pushes $R_m$ high,
so a $K_2$ of order 1 is desirable.
We conclude that (\ref{eq:Kweak}) is not sufficient for our model. 
To
surmount this problem, we assume that the K$\ddot{\rm a}$hler
potential develops strong coupling contributions 
\cite{BDi1}, \cite{BDi2}, \cite{BDi3}, just as
the superpotential does. The origin of these strong coupling
contributions can be the confining gauge group of the hidden
sector or other strong coupling phenomena at higher energy.   
Here, we will assume a K$\ddot{\rm a}$hler potential of the form:
$K^{\rm tot}(y)=K(y)+K^{\rm np}(y)$, where
\be K^{\rm np}(y)=ky^pe^{-ry^q}.\ee       
In the above, $p$, $q$ and $r$ are unknown numbers ($r$, $q$ $>0$). 
The constant $k$ can be fixed from the equation 
$(\xi /M)^2=4\pi ^2\delta_{GS}K^{\rm tot}_1(y_0)\label{eq:kxi}$,
where $\xi$ is the scale at which $X$ breaks. 
Of course, the
form of $K^{np}(y)$ could be more complicated or simply different, but in
any case, we will give arguments that indicate that it has to be some
kind of exponential function. Having $K^{\rm tot}(y)$ as our starting point,
we will try to answer the question, if it is possible 
to stabilize the dilaton in a similar fashion as with just $K(y)$ and in
addition to force $R_m$ to zero. This will have to involve in our
scheme a $K^{\rm tot}_2(y_0)$ of order of 1. We can express the above requirements
with the following equations:
\be K^{\rm tot}_2(s_0)\simeq 0,\label{eq:Kc1}\ee 
\be K^{\rm tot}_2(y_0)\simeq 1,\label{eq:Kc2}\ee  
\be R_m\sim 0: \;\;\;\; 
K^{\rm tot}_1(y_0)K^{\rm tot}_3(y_0)\simeq K^{\rm tot}_2(y_0)^2.\label{eq:Kc3}\ee
These are three constraints for three unknowns to be
determined. Unfortunately, the equations are very nonlinear so we are
not guaranteed a unique solution (not even a solution). In equations 
\ref{eq:Kc1} and \ref{eq:Kc2}, we already see the seed of the
reason for which we advocated that $K^{\rm np}(y)$ is some exponential
function. In order for $K^{\rm tot}_2(y)$ to increase from its value
$0$ at $y=s_0$ to $1$ at $y=s_0-\alpha$ with $\alpha$ very small, it
most probably has to be an exponential function.
We next give the derivatives of the K$\ddot{\rm a}$hler
potential:
\be K^{\rm tot}_1(y)=-{1\over y}+{s_0\over y^2}-{{4s_0^2+b}\over
{12y^3}}+\bigl({p\over y}-rqy^{q-1}\bigr)K^{\rm np}(y), \label{eq:conmin1}\ee
\begin{eqnarray} K^{\rm tot}_2(y)={1\over y^4}\Biggl\{(y-s_0)^2+{b\over
4}+(rqy^q)^2\Biggl[\biggl(y-({p\over {rqy^{q-1}}}-\sqrt{{p\over
{(rqy^{q-1})^2}}+{{(q-1)}\over {rq}y^{q-2}})}\biggr)\cdot \cr    
\biggl(y-({p\over {rqy^{q-1}}}+\sqrt{{p\over
{(rqy^{q-1})^2}}+{{(q-1)}\over {rq}y^{q-2}})}\biggr)\Biggr]K^{\rm
np}(y)\Biggr\},\label{eq:conmin2}\end{eqnarray} 
\begin{eqnarray} K^{\rm tot}_3(y)=-{2\over y^3}+{6s_0\over
{y^4}}-{{4s_0^2+b}\over {y^5}}+\bigl({{p\over y}-rqy^{q-1}}\bigr)K^{\rm
np}_2(y)- \cr 
2\bigl({p\over {y^2}}+rq(q-1)y^{q-2}\bigr)K^{\rm np}_1(y) 
+\bigl({2p\over {y^2}}-rq(q-1)(q-2)y^{q-2}\bigr)K^{\rm np}(y). 
\label{eq:conmin3}\end{eqnarray}
For the constant $k$, we obtain:
\be k={{{{\lambda_c^2}\over {4\pi^2\delta_{GS}}}+{1\over y_0^3}\bigl( 
y_0^2-s_0y_0+{{4s_0^2+b}\over 12}\bigr)}\over {y_0^p({p\over
y_0}-rqy_0^{q-1})e^{-ry_0^q}}
}\label{eq:k},\ee 
where we have set $\xi/M\sim \lambda_c$.  
Defining $x_y\equiv rqy_0^q$ and $x_s\equiv rqs_0^q\simeq x_y$, after a
considerable amount of algebra and always keeping only the dominant
contributions, we find that provided that $\alpha $ is
a small number, equations \ref{eq:Kc1}-\ref{eq:Kc3} reduce to 
\be x_s=p+{1\over 2}(q-1)\pm {1\over
2}\sqrt{(q-1)^2+4qp}\label{eq:sol1},\ee
\be x_y{{2{\hat \lambda} \alpha}\over {y_0^2}}=1\label{eq:sol2},\ee
\be {\lambda_c ^2\over {4\pi^2\delta_{GS}}} {{(-2q{\hat \lambda})}\over
{y_0^2}}x_y=1 \label{eq:sol3}\ee
respectively, where ${\hat \lambda}={\lambda_c ^2\over {4\pi^2\delta_{GS}}}+{b\over
{12y_0^3}}+{1\over {3y_0}}-{\alpha\over {3y_0^2}}+{\alpha^2\over
{3y_0^3}}\simeq \lambda_c $. If we recall \cite{ILR} that
$y_0$, $\delta_{GS}$ and $\lambda_c $ are calculable numbers, we
see that the parameters $p,q$ and $r$ can be derived from
\ref{eq:sol1} to \ref{eq:sol3}, which shows that there exist $p,q$ and $r$
that satisfy \ref{eq:Kc1}-\ref{eq:Kc3}.
However, since we do not have any
physical intuition at the moment what are the expected values of these
parameters, we will not give specific examples. 
The lesson from this analysis is that we need to have 
$K^{tot}_2(y_0)=1$ to suppress fcnc and that this is possible only if
the K$\ddot{\rm a}$hler potential has some additional contributions,
probably due to strong coupling physics. In a future work we will
investigate the different possibilities and constraint them by looking
at cosmological issues \footnote{An alternative, interesting form that
will be examined from this point of view in a future work will be 
$K^{np}(y)=c_1+c_2\int_0^y{e^{-4\pi^2(t-s_0)^2}}dt$, which along with
$K^0(y)=-ln(2y)-{(s_{-1}+s_0)\over {2y}}+{{2/3s_{-1}s_0}\over
{(2y)^2}}$ ($s_{-1}$, $c_1$ and $c_2$ are constants), 
stabilizes the dilaton at $s_0$.}. 
Fortunately, we do not need to have an explicit form for $K^{tot}$
in order to make predictions about low energy
physics, as long as we require that whatever form it has, it is such
that $K^{tot}_2(y_0)=1$.  
\par The superpotential for a 
QCD like
hidden sector (that develops gaugino
condensates) with gauge group $SU(N_c)$ and $N_f$ families ($N_f<N_c$) is
\be W={1\over 2}Mt^2\Bigl({\theta _0\over M}\Bigr)^{p_0}
\Bigl({\theta _1\over M}\Bigr)^{p_1}\Bigl({\theta _2\over
M}\Bigr)^{p_2}+\bigl({{N_c}-N_f}\bigr)
\Bigl({2\Lambda ^{{\beta_0}\over 2}\over t^2}\Bigr)^{1\over 
{{N_c}-N_f}}\label{eq:w} \label{eq:W},\ee 
where $t$ is the ``quark'' condensate and 
${\beta_0}=2(3N_c-N_f)$ is the one loop $\beta$
function. Also, $\Lambda=Me^{-8\pi ^2k _h(2S)/{\beta_0}}$,
with $k_h$ the Kac-Moody level of the hidden group 
and \cite{BD}, 
\be {\hat m}=M\Bigl({<\theta _0>\over M}\Bigr)^{p_0}
\Bigl({<\theta _1>\over M}\Bigr)^{p_1}\Bigl({<\theta _2>\over
M}\Bigr)^{p_2}\;\;
{\rm and}\;\;\; \epsilon =\Bigl({\Lambda \over
\xi}\Bigr)^{{\beta_0}\over {2N_c}}\Bigl({\xi
\over {\hat m}}\Bigr)^{1-{N_f\over N_c}}.\ee
The $F$-term contribution to the
soft masses then is \cite{dilaton}, \cite{I1}
\be m_{0}^2=\Bigl[{2\sqrt{2}\over {\sqrt{K^{\rm
tot}_2(y_0)}}}(\epsilon{\hat
m})(8\pi^2k_h)\lambda_c^2\Bigr]^2\;,\;\;\; 
m_{1/2}\simeq m_0\simeq {\rm a}_0\label{eq:mass}\ee 
and the K$\ddot{\rm a}$hler potential dependence only comes through 
$K_2^{\rm tot}$ at the minimum, which as we argued, is equal to 1. 
\par We are now ready for an explicit numerical example. 
We first recall the values of the 
hidden sector parameters (which are the relevant ones for our case too) 
found in \cite{I1}: $N_c=5$, $N_f=3$, 
$\delta_{GS}=-0.113$,
$p_0=p_1=p_2=6$, $k_h=1$, $<\theta _i>/M=0.222\simeq
\lambda_c$. Second, we recall that in the same model, the scale 
at which the gauge couplings unify is  
$M_{GUT}\simeq 4\times 10^{16}$
GeV and the vacuum expectation value of the dilaton,
(the one loop unified gauge coupling at $M_{GUT}$) is
$y_0=1/g^2(M_{GUT})=1.429$, 
Third, we will
choose the sign of the $\mu$ term to be positive, since the negative
choice tends to give problems with vacuum stability. Given the above,
we end up having only two free parameters:
$\tan{\beta}$ and the cut off scale $M$, 
which need not be necessarily be $M_{GUT}$.
In addition, if we       
remind ourselves that in this model we have $m_b/m_t\sim \lambda_c^3$,
we conclude that $\tan{\beta}$ can not be larger than about 5 if
we do not want to introduce unnaturally small numerical coefficients
in front of the Yukawa couplings.
Having $m_0$, $m_{1/2}$, ${\rm a}_0$ (from \ref{eq:mass}), 
$\tan{\beta}$ and $sgn(\mu)$, we
can make specific 
low energy predictions if we run the parameters via the RGE
equations. 
We present a table with the prediction of the Higgs mass 
for different values of $\tan{\beta}$ and $M$.  
\vskip 0.3cm
\begin{center}
\begin{table}
\caption{Higgs mass versus $\tan{\beta}$ for different values of $M$. 
The second column is for $M\simeq M_{GUT}=4\cdot 10^{16}$ GeV, 
the third for $M=8\cdot 10^{16}$ GeV 
and the fourth for $M=1.2\cdot 10^{17}$ GeV.}
\vskip 0.5 cm
\begin{center}
\begin{tabular}{|c||c|c|c|}
\hline
${\tan{\beta}}$&${h_0}\; ({\rm GeV})$
&${h_0}\; ({\rm GeV})$
&${h_0}\; ({\rm GeV})$\\
\hline
$2$&$79.0$&$90.0$&$95.0$\\ 
$3$&$80.2$&$92.0$&$97.0$\\ 
$4$&$91.5$&$103.0$&$107.0$\\ 
$5$&$99.6$&$109.5$&$112.0$\\ 
\hline
\end{tabular}
\end{center}
\end{table}
\end{center}
\vskip 0.3cm  
\par We close this section by giving arguments to show the fact
that supergravity effects will have negligible corrections
to this picture. The ratio $R_m$ including supergravity corrections 
can be written as \cite{BDi3}
\be R_m^{\rm SUGRA}={\bigl[-{K^{tot}_3\over 
K^{tot}_1}+({K^{tot}_2\over K^{tot}_1})^2+\Delta _2({K^{tot}_2\over
K^{tot}_1})^2-4\pi^2\delta_{GS}(\Delta_1+1)({K^{tot}_2\over K^{tot}_1})^3\bigr]
\over {2K^{tot}_2(1-4\pi ^2\delta_{GS}{K^{tot}_2\over K^{tot}_1}})}+{{\Delta_{3}}\over
K^{tot}_2}({K^{tot}_2\over K^{tot}_1})^2,\ee  
where (for $N_f=1$):
\be \Delta _1={{z^2+6nz-4n^2}\over {(z-2n)^2}}, \;\;\; 
\Delta _2={{2z^2+2nz}\over {(z-2n)^2}},\ee 
where $z=2N_c\lambda_c^2$
and $\Delta_{3}$ is given by \cite{BDi3} 
\be \Delta_{3}={1\over 2}K^{tot}_1({\tilde \delta_{GS}}-4{K^{tot}_1\over K^{tot}_2})-{N_c\over
n^2} {{\tilde \delta_{GS}}{K^{tot}_1}^2\over {4K^{tot}_2}}(-nK^{tot}_1+{1\over
4}(N_c+n){\tilde \delta_{GS}}K^{tot}_2)-{N_c\over n}{3{\tilde
\delta_{GS}}K^{tot}_1\over 4},\ee
where ${\tilde \delta_{GS}}=16\pi^2\delta_{GS}$.
We can easily check that
$\Delta _1\simeq -1$, $\Delta_2\simeq 0$ and $\Delta _3\simeq
0$ which demonstrates that the global supersymmetry limit was
sufficiently good for our numerical predictions, since only equation
\ref{eq:Kc3} is modified slightly from $R_m\sim 0$, 
to $R_m^{\rm SUGRA}\sim 0$. 

\section{Conclusions}

We have shown that in models with $U(1)$ family symmetries the 
alignment mechanism can be implemented via the appearance of
holomorphic zeros in the superpotential -which tend to destabilize the
vacuum by opening flat directions. On the alternative, there
can be constructed models of fermion masses with minimal number
of holomorphic zeros which in order to be viable from the fcnc point of view 
are complemented by a hidden sector and a dilaton dominated supersymmetry breaking
mechanism. Such a mechanism that at the same time stabilizes the
dilaton and suppresses the $D$-term contributions to the soft masses
is presented. An explicit numerical example demonstrated that the
model gives a small Higgs mass, very close in some cases to its
current central value. Even though it is true that a low Higgs mass is a generic
feature of supersymmetric models, we think that it is
remarkable that a model that has such a few free parameters, gives
predictions in the expected range.   

\section{Acknowledgements}

I would like to thank P. Ramond for discussions and his useful
comments
and R. Haas for his help with SPYTHIA. 
NI is supported in part by the 
United States Department of Energy under grant DE-FG02-97ER41029.


\end{document}